\documentclass{aa}  

\usepackage{graphicx}
\usepackage{txfonts}
\usepackage{hyperref}
\hypersetup{colorlinks,linkcolor={blue},citecolor={blue},urlcolor={blue}}  
\usepackage{orcidlink}

\usepackage{csquotes}

\begin{document} 

   \title{Diffuse neutrino flux from relativistic reconnection in AGN coronae}
   \titlerunning{Neutrinos from AGN coronae}
   \author{D. Karavola
          \inst{1} \and M. Petropoulou \inst{1, 2} \and
          D. F.~G.~Fiorillo\inst{3} \and A. Georgakakis\inst{4} \and L. Comisso\inst{5} \and L. Sironi\inst{5,6} 
          }

   \institute{Department of Physics, National and Kapodistrian University of Athens, University Campus Zografos, GR~15784, Athens, Greece\label{1} \and Institute of Accelerating Systems \& Applications, University Campus Zografos, GR 15784, Athens, Greece\label{2}
   \and Istituto Nazionale di Fisica Nucleare (INFN), Sezione di Napoli,
    Complesso Universitario di Monte Sant’Angelo, Via Cintia, 80126 Napoli, Italy\label{3} \and Institute for Astronomy \& Astrophysics, National Observatory of Athens, V. Paulou \& I. Metaxa, 11532, Greece\label{4}\and Department of Astronomy and Columbia Astrophysics Laboratory, Columbia University, New York, NY 10027, USA\label{5} \and Center for Computational Astrophysics, Flatiron Institute, New York, NY, 10010, USA\label{6}\\ \email{dkaravola@phys.uoa.gr, mpetropo@phys.uoa.gr, damiano.fiorillo@desy.de}}

   \date{Received ; accepted }

  \abstract
   {IceCube observations point to active galactic nuclei (AGNs) as promising contributors to the observed astrophysical neutrino flux. Close to the central black hole, protons can be accelerated through magnetic reconnection to very high energies and subsequently interact with abundant X-ray photons in the source, leading to neutrino production.}
   {We investigate whether the diffuse neutrino flux observed by IceCube can originate, via proton acceleration, in reconnection-powered coronae of non-jetted AGNs.}
    {We created a library of neutrino spectral templates over a large grid of values for the three key model parameters: the proton plasma magnetization of the corona $\sigma_{\rm p}$, the X-ray coronal luminosity, and the black hole mass. Synchrotron cooling of pions and muons plays a significant role due to the large coronal magnetic fields. To infer the diffuse neutrino flux, we coupled the single-source model with a mock AGN catalog consistent with the observed X-ray and mid-infrared AGN samples at redshifts  $z=0-4$.}
   {The coronal emission satisfactorily explains the most recent IceCube measurements of the diffuse neutrino flux up to energies of $\sim 1$~PeV, provided that $\sim$10\% of the AGN coronae have $\sigma_{\rm p} \sim 10^5$, 
   while the rest are distributed over a range of lower magnetizations. Coronal emission is suppressed at higher energies by pion and muon cooling so that another population is required, with jetted AGNs
   being strong candidates.}
  {}

   \maketitle
%

\section{Introduction}
IceCube has measured a diffuse flux of astrophysical neutrinos with a spectrum roughly consistent with a power law, $\Phi_\nu \propto E_\nu^{-\gamma}$, over the range $\sim 10$~TeV–10~PeV, with $\gamma = 2.4{-}2.9$ depending on the specific analysis considered \citep[e.g.,][]{Abbasi2021, Abbasi2022}. Since its discovery in 2013 \citep{IceCube2013}, improved statistics and analysis techniques have refined this picture. The latest results extend the spectrum down to $\sim 1$~TeV and have revealed evidence of a low-energy break near 10~TeV \citep{basu_measurement_2025}, indicating that the diffuse neutrino spectrum deviates from a simple power law. The origin of this unresolved emission remains unknown.

In addition to the measurement of the diffuse flux, searches for neutrino point sources have been performed over the past decade.
To date, only two astrophysical neutrino sources have been identified at a high confidence level, and both are associated with active galaxies of different types: the blazar TXS~0506+065 \citep{collaboration_neutrino_2018, padovani_dissecting_2018} and the nearby Seyfert II NGC~1068, which stands as the ``hottest'' spot in the neutrino sky \citep{abbasi_search_2024, abbasi_evidence_2025}. 

The detection of neutrinos from NGC 1068, in combination with the absence of very high energy $\gamma$-rays, points toward neutrino production in a compact $\gamma$-ray opaque region, the corona, which is located in the vicinity of the central black hole. Three main theoretical scenarios have been proposed to model proton acceleration in active galactic nuclei (AGN) cores, namely, diffusive shock acceleration at shock waves~\citep[e.g.,][]{stecker_high-energy_1991, inoue_origin_2020}, stochastic acceleration in magnetohydrodynamic turbulence \citep[e.g.,][]{murase_hidden_2020, fiorillo_magnetized_2024, lemoine_neutrinos_2025, saurenhaus_constraining_2026}, and systematic acceleration in magnetic reconnection \citep[e.g.,][]{fiorillo_reconnection_2024, karavola_neutrino_2025}. In a magnetically powered corona, the last two mechanisms are generally considered to accelerate protons to nonthermal energies. 
In turbulence-based models, neutrinos are produced by p$\gamma$ and p-p interactions in a coronal region of $\sim10-100$ gravitational radii. Based on this scenario, \cite{murase_hidden_2020} and \cite{padovani_neutrino_2024} determined the diffuse neutrino spectrum expected from these turbulent coronae according to the model of \cite{murase_hidden_2020}. More recently, \cite{fiorillo_contribution_2025} calculated the neutrino diffuse flux according to the strongly magnetized turbulence scenario of \cite{fiorillo_magnetized_2024}. The general consensus is that turbulent AGN coronae may account for the $\sim$1-10~TeV diffuse neutrino emission but not the higher energy tail, due to strong Bethe-Heitler proton cooling.

In this paper, we calculate the diffuse neutrino flux arising from non-jetted AGNs based on the magnetic reconnection coronal model introduced by \cite{fiorillo_reconnection_2024,karavola_neutrino_2025}. In this model, protons are accelerated in magnetospheric current sheets and on average obtain an energy of $E_{\rm p} \sim \sigma_{\rm p} m_{\rm p} c^2$ \citep{com2024, sironi_25, hakobyan_reconnection-driven_2025}, where $\sigma_{\rm p}=B^2/(4\pi n_{\rm p} m_{\rm p} c^2) \gg 1$ is the proton plasma magnetization; here, $B$ is the magnetic field strength and $n_{\rm p}$ is the number density of non-relativistic protons in the corona. Relativistic protons interact with coronal X-ray photons, producing pions. Neutrinos with a typical energy of $E_{\nu} \sim E_{\rm p}/20$ are subsequently produced through the decays of pions and muons. Our model is characterized by three key parameters: the proton plasma magnetization ($\sigma_{\rm p}$), the coronal X-ray luminosity ($L_{\rm X}$), and the coronal radius ($R$; or equivalently the black hole mass, $M$). 
We coupled this single-source model with an AGN mock catalog \citep{georgakakis_forward_2020} to compute the cumulative neutrino flux from AGN coronae up to redshift $z = 4$. We show that the observed diffuse neutrino flux in the 1 TeV--1 PeV range can be naturally reproduced if $\sim$10\% of the AGN coronae have $\sigma_{\rm p} \sim 10^5$, while the rest are distributed over a range of lower magnetizations.\footnote{To date, $\sigma_{\rm p}$ has not been directly constrained by observations and is therefore usually inferred through modeling.}

\section{Diffuse neutrino flux calculation}
In this section we outline the corona model and introduce its key parameters. We then present the AGN mock catalog and describe the diffuse neutrino flux calculation.

\subsection{Corona model} \label{sec:corona}
We assumed that neutrinos in non-jetted AGNs are produced within a compact, highly magnetized region close to the black hole coinciding with the site of nonthermal X-ray emission commonly referred to as the corona. We briefly outline the model introduced in \cite{fiorillo_reconnection_2024} to explain the IceCube observations of NGC~1068 and explored more generally for AGN coronae in \cite{karavola_neutrino_2025}. In this model, the powerhouse of coronal emission is magnetic reconnection in current sheets that form within the magnetospheric region of an accreting black hole. The current sheet is described by a cuboid with dimensions $L\times L \times \beta_{\rm rec}L$, with $L= \bar R r_{\rm g}$, $\beta_{\rm rec} \sim 0.1$ being the reconnection rate and  $r_{\rm g} = GM/c^2$. For the scope of the numerical calculations, we describe the corona as a spherical environment with volume $\beta_{\rm rec} L^3$. Under this description, one can define an effective coronal radius: $R=\left[3 \beta_{\rm rec}/(4 \pi) \right]^{1/3} L \sim 0.29 L$. In this work, we adopted $ \bar R = 12$, motivated by global numerical simulations \citep{ripperda_black_2022}. 

A fraction, $\eta_{\rm X} \sim 0.5$, of the dissipated magnetic energy is radiated as X-rays, with a bolometric luminosity (0.1–100 keV) given by $L_{\rm X}= \eta_{\rm  X} c \beta_{\rm rec} B^2 R^2$, where $B$ is the strength of the reconnecting magnetic field, and $R$ is the characteristic size of the corona. A fraction, $\eta_{\rm p} \sim 0.1$, of the dissipated energy is transferred to relativistic protons with luminosity: $ L_{\rm p} = (\eta_{\rm p}/\eta_{\rm X}) L_{\rm X}$. We discuss the specific choices of $\eta_{\rm X}$ and $\eta_{\rm p}$ in Sec.~\ref{sec:disc}. The distribution function of accelerated protons is a broken power law, $dN/d\gamma \propto \gamma^{-1}$ for $1 < \gamma < \gamma_{\rm br}$ or $\gamma^{-s_{\rm p}}$ otherwise, with the post-break slope ($s_{\rm p}$) depending on the guide-field\footnote{This is the non-reconnecting magnetic field component.} strength \citep{com2024}. Here we adopted $s_{\rm p}=3$, which corresponds to strong guide-field reconnection.

The break Lorentz factor is determined by the proton plasma magnetization, $\gamma_{\rm br} \sim \sigma_{\rm p}= B^2/(4 \pi n_{\rm p} m_{\rm p} c^2$ \citep{com2024, hakobyan_reconnection-driven_2025}, where $n_{\rm p}$ is the plasma density of (non-relativistic) protons in the corona (upstream of the reconnection layer). We envisioned that the reconnection layer is formed in a region distinct from the dense accretion disk whenever steady accretion is temporarily halted due to the accumulation of magnetic flux. As a result, the proton density in this region is expected to be extremely low, allowing $\sigma_{\rm p}$ to reach high values.\footnote{Transient baryon-poor regions, extended over a large range of polar angles, have been found to be formed in the magnetosphere of magnetically arrested disks \citep{chow_baryon_2026}.}

In our model, high-energy neutrinos are produced through interactions of relativistic protons with coronal X-rays. \cite{karavola_neutrino_2025} demonstrated that the resulting neutrino emission is controlled by two parameters: the proton magnetization ($\sigma_{\rm p}$) and the AGN X-ray Eddington ratio, defined as $\lambda_{\rm X, Edd}= L_{\rm X}/L_{\rm Edd}$, where $L_{\rm Edd} = 4 \pi G M m_{\rm p} c /\sigma_T$ is the Eddington luminosity of the accreting black hole. 

We also accounted for the synchrotron cooling of muons and pions (as described in Appendix~\ref{appA}), which becomes important in the strong magnetic fields considered here, $B \sim 0.1 - 10^3$~kG. Synchrotron cooling introduces characteristic break energies to the neutrino spectrum, thus breaking the self-similarity of spectral shapes with respect to $\lambda_{\rm X, Edd}$ -- see Appendix \ref{appA} --  and softens the neutrino spectra, making the imprint of the proton slope less prominent, as shown in Appendix \ref{App:Slopes}.

We created neutrino spectral templates using the leptohadronic code ATHE$\nu$A \citep{2012A&A...546A.120D} for a parameter grid spanning $\log_{10}L_{\rm X} = 42 - 47$, $\log_{10}(\sigma_{\rm p})=3-5$, and $\log_{10}(R) = 12.3 - 14.3$ in integer steps. These ranges were motivated by the AGN mock catalogs described in the next section.

\subsection{AGN mock catalog} \label{sec:mock}
To calculate the neutrino emission from a population of AGNs, the X-ray luminosity and the black hole mass of each source are required. For this purpose, we used the simulated AGN sample of \cite{georgakakis_forward_2020}, which provides the 2–10 keV X-ray luminosity\footnote{This is extrapolated to the 0.1$-$100 keV range of the corona model, assuming a photon index $s_{\rm x}$ of two. We discuss the impact of $s_{\rm x}$ in App. \ref{App:Slopes}.} and the stellar mass ($M_*$) of the host galaxy. This $M_*$ is assigned to a black hole mass according to the redshift-independent relation $M =2 \cdot 10^{-3} M_*$ \citep{marconi_relation_2003}, and it translates to a current sheet dimension, $L = 12 \, r_{\rm g}(M)$. The catalog was constructed using the X-ray luminosity function of \cite{2014ApJ...786..104U} as described in Appendix~\ref{appB}.

\subsection{Diffuse flux calculation}
We denote the template neutrino luminosity spectra on the discrete grid of proton magnetization ($\sigma_{\rm p}$), the X-ray luminosity ($L_{\rm X}$), and the coronal radius ($R$) as $\mathcal{L}_{(a,b,c)} \equiv E_{\rm em}L(E_{\rm em}; \sigma_{\rm p}^{(a)}, L^{(b)}_{\rm X}, R^{(c)})$, where $E_{\rm em}$ is the neutrino energy in the rest frame of the galaxy, $a=1,...,3$, $b = 1, ..., 6$, and $c=1, ...,3$. The neutrino flux spectrum of a simulated AGN corona with $\sigma_{\rm p}^{(i)}$, $L^{(i)}_{\rm X}$, and $R^{(i)}$ at redshift $z_i$ was then derived as
\begin{eqnarray}
E_{\rm obs} F^{(i)}(E_{\rm obs})  =  \frac{1}{4\pi D_{\rm L}^2(z_i)} E_{\rm em} L(E_{\rm em}; \sigma_{\rm p}^{(i)}, L^{(i)}_{\rm X}, R^{(i)}) ,
\end{eqnarray}
where $E_{\rm obs}=E_{\rm em}/(1+z_i)$, $D_{\rm L}$ is the luminosity distance\footnote{We adopted a cosmological model with $H_0=69.32~{\rm km/(Mpc~s)}$, $\Omega_{\rm m}=0.29$ \citep{planck_collaboration_planck_2020}.} and $\mathcal{I}[\mathcal{L}_{(a,b,c)}]$ is the interpolation operator. We performed a trilinear interpolation of the template spectra in logarithmic space, i.e., we interpolated $\log_{10}(E_{\rm em} L(E_{\rm em}))$ as a function of $\log_{10}L_{\rm X}$, $\log_{10}R$ and $\log_{10}(\sigma_{\rm p})$, on a fixed logarithmic energy grid in the source rest frame ($\log_{10}E_{\rm em}$). 

\cite{fiorillo_reconnection_2024} and \cite{karavola_neutrino_2025} showed that a value of $\sigma_{\rm p} \sim 10^5$ was required to reproduce the IceCube spectrum of NGC~1068, which remains the brightest neutrino source observed to this day. Motivated by the aforementioned result, we initially assumed a common $\sigma_{\rm p}=10^5$ for all AGNs, which led to an overestimation of the diffuse flux by a factor of roughly eight at $E_\nu=10$~TeV (see the black line in Fig.~\ref{fig:slopes}). Therefore, sources with $\sigma_{\rm p} \sim 10^5$ should constitute $\sim 10\%$ of the AGN population to avoid overshooting the diffuse flux \citep{abbasi_evidence_2025}.

\begin{figure}
    \centering
    \includegraphics[width=0.95\linewidth]{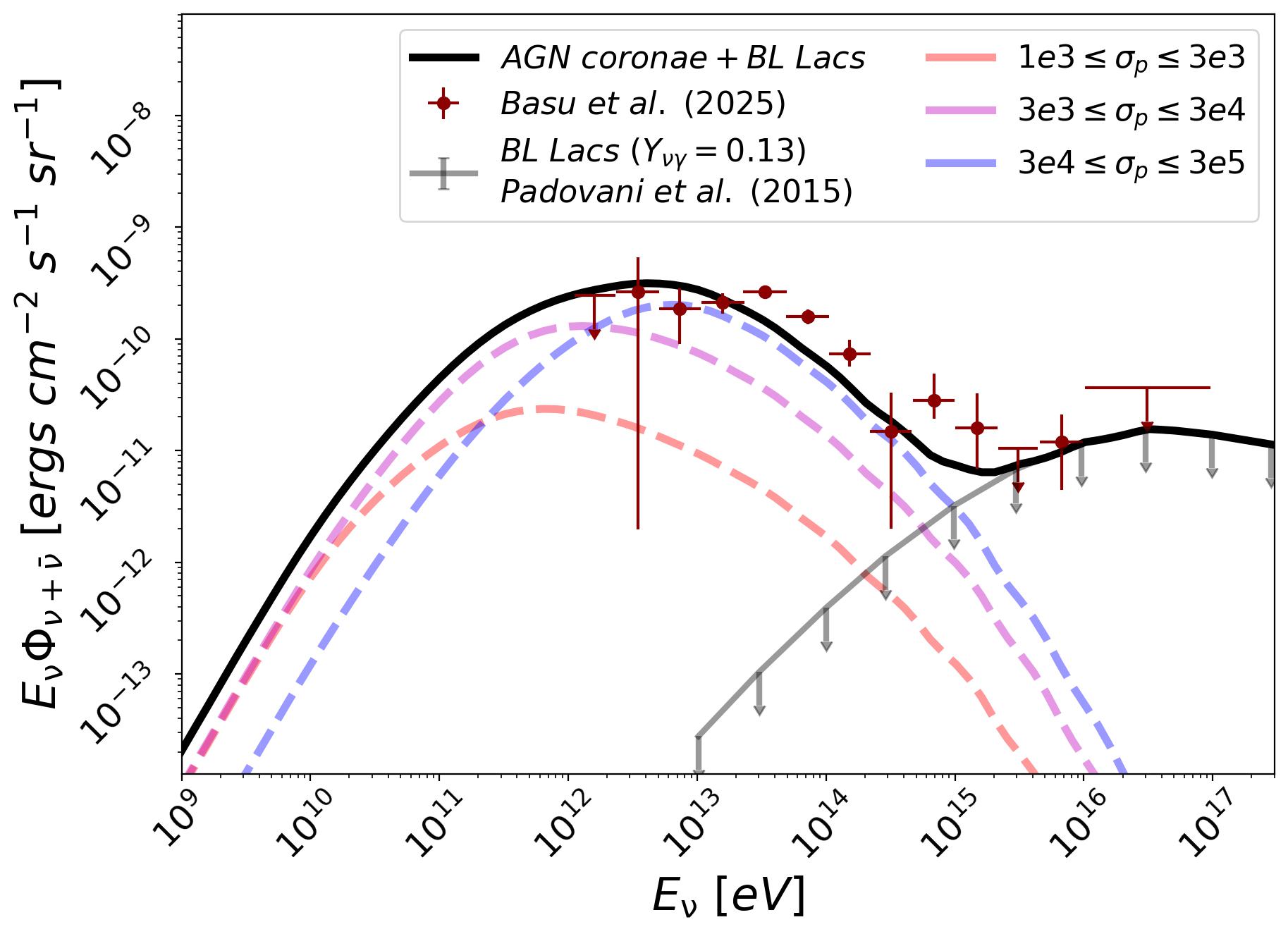}
    \caption{Diffuse all-flavor neutrino flux spectrum. Burgundy markers show the IceCube diffuse flux spectrum as obtained from the most recent analysis of \cite{basu_measurement_2025}. The solid black line shows the total flux. This comprises emission from AGN coronae predicted by our model (up to $\sim 10^{15}$ eV) and from jetted AGNs (BL Lac sources according to \cite{padovani_simplified_2015}) at higher energies. 
    The contribution of sources with different $\sigma_{\rm p}$ values is shown with dashed colored lines -- see the legend.}
    \label{fig:diffuse_neutrino_spec}
\end{figure}

Next, we assigned a random value ($\sigma_{\rm p}$) to each source sampled from a power-law probability distribution, $P(\sigma_{\rm p}) \propto \sigma_{\rm p}^{-n}$ with $n>1$ and $\log_{10}(\sigma_{\rm p}) \in [3, 5]$. As a result, we argue that sources with such a high proton magnetization should be rare in the population. The results shown in Sec. \ref{sec:diffuse} were obtained for $P(\sigma_{\rm p}) \propto \sigma_{\rm p}^{-2}$. Predictions made for varying $n$ and different probability distributions are presented in Appendix \ref{appC}. Finally, the neutrino energy flux from the whole sky (per steradian) using the two mock catalogs was computed as
\begin{equation}
E_{\rm obs}^2 \Phi_{\nu + \bar{\nu}}(E_{\rm obs}) = (4\pi)^{-1}  \sum_{j}\frac{4\pi} {\Omega_{\rm j}} \sum_{i=1}^{N_{\rm j}} E_{\rm obs} F^{(i)}(E_{\rm obs}), 
\end{equation}
where $j=\{1,2\}$, while $\Omega_j=\{50 \,~(\pi/180)^2, ~2\cdot 10^4 \, (\pi/180)^2 \}$,  and $N_j=\{1.2 \cdot 10^6, ~7.1 \cdot 10^5\}$ are the solid angles and number of sources in the $0.2 \leq z\leq 4$ and $z \leq 0.2$ sub-catalogs, respectively.

\section{Results} \label{sec:diffuse}
Figure \ref{fig:diffuse_neutrino_spec} shows the results for the diffuse all-flavor neutrino flux for the whole sky, marked with the solid black line, up to $E_{\nu}=10^{17}$~eV. The solid gray line represents the maximum contribution of BL Lacs allowed by IceCube upper limits (beyond PeV energies) based on the model of \cite{padovani_simplified_2015}. Moreover, with burgundy markers we show the IceCube data as given by the most recent analysis of \citep{basu_measurement_2025}. Dashed lines of different colors indicate the contribution of sources with varying $\sigma_{\rm p}$ values. We find that the total diffuse neutrino spectrum is not particularly sensitive to the choice of $P(\sigma_{\rm p})$ but to the fraction of the sources with $5 \cdot 10^4 \leq \sigma_{\rm p} \leq 5 \cdot 10^5$. In particular, we are able to describe the IceCube observations if the aforementioned population constitutes 10\% of the total non-jetted AGNs.

Our model is able to describe the observational data up to $E_{\nu} \sim 10^{15}$ eV. Although coronae with $\sigma_{\rm p}=10^5$ are the least common in the sample, their contribution to the total flux is the largest. This results from the dependence of the neutrino production efficiency on $\sigma_{\rm p}$: As the latter increases, the neutrino production efficiency also increases until saturation is achieved, as discussed in \cite{karavola_neutrino_2025}. 
Suppression of the neutrino flux at higher energies ($E_{\nu} \gtrsim 10^{15}$~eV) due to muon and pion cooling leaves room for other source classes to contribute. For instance, gray solid line illustrates the maximum expected contribution from jetted AGNs -- particularly BL Lac objects -- based on the model of \cite{padovani_simplified_2015}. The combined spectrum from both non-jetted and jetted AGNs provides a very good description of the diffuse flux across four orders of magnitude in energy.

\section{Discussion and conclusions}\label{sec:disc}
 
We calculated the diffuse neutrino flux according to our reconnection-powered AGN corona model. We showed that AGN coronae could account for the majority of the diffuse neutrino background up to energies of $10^{15}$eV, provided that approximately $10 \%$ of the population has coronae with proton magnetization in the range $5 \cdot 10^4 \leq \sigma_{\rm p} \leq 5 \cdot 10^5$. Moreover, we found that galaxies at $0.5 \leq z \leq 3.0$ with $10^{44}\,  {\rm erg~s}^{-1} \leq L_{\rm X} \leq 10^{46} \, {\rm erg~s}^{-1}$ are the ones that mostly contribute to the diffuse neutrino flux at energies $ 1~{\rm TeV}\leq E_{\nu} \leq 1~{\rm PeV}$ -- see Appendix \ref{appB}. At higher energies the coronal neutrino spectrum is suppressed by pion and muon synchrotron cooling, leaving room for other source classes to contribute to multi-PeV energies.

In this study, we adopted a post-break slope of $s_{\rm p}=3$, which was motivated by the results of \cite{fiorillo_reconnection_2024}, who showed that this value is required to reproduce the neutrino spectrum of NGC~1068. Smaller values of $s_{\rm p}$ lead to a harder neutrino spectrum at energies $\gtrsim \sigma_{\rm p} m_{\rm p} c^2/20$. In particular, for $s_{\rm p}=2$, which would be expected from reconnection in the limit of a vanishing guide field, the neutrino spectral peak becomes insensitive to $\sigma_{\rm p}$ and is controlled by the maximum proton energy. For these reasons, zero-guide field reconnection in NGC 1068 has been disfavored by previous works \citep{fiorillo_reconnection_2024}. However, the effects of  pion and muon cooling were not accounted for in previous works. In Appendix~\ref{App:Slopes}, we relax the constraint on $s_{\rm p}$ and demonstrate that once pion and muon cooling losses are included, the high-energy neutrino slope obtained for $s_{\rm p}=2$ closely resembles the result of $s_{\rm p}=3$ in the absence of heavy-particle cooling (as assumed by \cite{fiorillo_reconnection_2024}). As a result, reconnection in the vanishing guide-field regime appears perfectly consistent with the neutrino spectrum of NGC 1068, as long as pion and muon cooling losses are properly included.

In our coronal model, we assumed that pairs accelerated during magnetic reconnection transfer their energy to X-ray photons. Both 2D and 3D PIC simulations \citep{sironi15, Werner_2017} have shown that the fraction $\eta_{\rm X}$ of electromagnetic energy dissipated into pairs decreases with increasing guide-field strength ($B_{\rm g}$), ranging from $\sim 0.5$ to $\sim0.1$ for vanishing and strong guide fields, respectively. Because the proton energy fraction, $\eta_{\rm p}$, shows a similar trend, the ratio $\eta_{\rm p}/\eta_{\rm X}$, which is crucial for our model, is rather insensitive to $B_{\rm g}$. Instead, its value depends on the pair loading of the plasma. 
In particular, \citet{petropoulou_relativistic_2019} found that $\eta_{\rm p}/\eta_{\rm X} \sim 0.1$ in pair-dominated plasmas, increasing toward unity in electron-proton plasmas. Motivated by these results and given the significant pair enrichment in our model \citep{karavola_neutrino_2025}, we adopted $\eta_{\rm X}=0.5$ and $\eta_{\rm p}/\eta_{\rm X}=0.2$.

In the coronal model used in this study, reconnecting current sheets form not within the accretion disk itself but in the magnetospheric region around the black hole. Plasma from the disk can be intermittently funneled into this region if accretion is regulated by the magnetic flux that threads the inner disk \citep{ripperda_black_2022}. Only a small fraction, $\zeta$, of the density of the particles in the disk is expected to enter the magnetosphere, with $\zeta$ not yet quantified in simulations. The proton magnetization scales as $\sigma_{\rm p} \propto B^2 / (\zeta n_{\rm p, disk})$,  and it is independent of both $L_{\rm X}$ and $M$ since $B^2\propto L_{\rm X}/M^2$ and $n_{\rm p, disk} \propto L_{\rm X} / M^2$ \citep[see Eqs.~E1 and E2 in][]{fiorillo_reconnection_2024}. Consequently, the distribution in $\sigma_{\rm p}$ needed to explain why the diffuse flux implies a corresponding dispersion in the fraction $\zeta$ across AGN coronae. Although the underlying physical driver of this dispersion is still uncertain, it may indicate that the formation and dynamics of magnetospheric current sheets differ among AGNs with varying coronal conditions.

Most studies adopt the hypothesis that the neutrino luminosity of the corona scales linearly with its X-ray luminosity \citep[e.g.,][]{padovani_neutrino_2024}. A key feature of our model is that the neutrino--X-ray luminosity scaling depends on the coronal photon compactness, $L_{\rm X}/R$. For high compactness, the corona is optically thick to p$\gamma$ interactions and $L_{\nu} \propto L_{\rm X}$, while $L_{\nu} \propto L_{\rm X}^2$ otherwise \citep{karavola_neutrino_2025}. This feature allowed us to loosen the commonly adopted assumption that all non-jetted AGNs follow a behavior similar to the neutrino-bright NGC~1068. As opposed to models that associate the corona with magnetized turbulence in the accretion flow \citep{murase_hidden_2020, fiorillo_magnetized_2024, lemoine_neutrinos_2025}, in our scenario neutrinos are produced due to p$\gamma$ interactions in the magnetospheric current sheets. In these environments, p-p collisions are not important, compared to p$\gamma$ processes, as was discussed in \cite{fiorillo_reconnection_2024}. This is due to the fact that the plasma in the upstream region is pair-dominated, and thus the number density of non-relativistic protons is significantly lower than the density of leptons and photons.

Given that our model can reproduce nearly the entire diffuse neutrino flux at energies 1-10 TeV, it is natural to ask what implications it has for the diffuse $\gamma$-ray background. Photons with energies $E_\gamma \gtrsim 10$~MeV are produced due to Bethe-Heitler and p$\gamma$ processes. However, these are attenuated due to the high opacity of the corona to $\gamma \gamma$ interactions with its X-ray photons. The photon energy flux at $E_\gamma \sim 1-10$~MeV in the corona is comparable to the peak neutrino energy flux \citep{karavola_neutrino_2025}. Therefore, the diffuse 1-10 MeV flux is expected to be comparable to the peak diffuse neutrino flux, which is a few times lower than the measured $\gamma$-ray flux in these energies (see figure 10 in \cite{ackermann_spectrum_2015}).     

Beyond the diffuse neutrino flux, our model also has important implications for individual AGN sources. In previous work we provided source-specific predictions \citep{karavola_neutrino_2025}, and a natural next step is to use the neutrino spectral templates developed here to perform a more robust statistical comparison with IceCube data. Such an analysis would allow us to infer plausible values of $\sigma_{\rm p}$ for AGNs with neutrino associations or to place meaningful upper limits on $\sigma_{\rm p}$  for non-detected sources. This is particularly relevant given that the proton magnetization of the magnetospheric environment of AGN remains one of the key unknown parameters of the model.

\begin{acknowledgements}
We thank the anonymous referee for their insightful comments that helped us clarify several points in the manuscript. D.K. thanks V.C. Karavolas for the useful conversations during the development of this paper.
DFGF is supported by the Alexander von Humboldt Foundation (Germany) and, when this
work was started, was supported by the Villum Fonden (Denmark) under Project No. 29388
and the European Union’s Horizon 2020 Research and Innovation Program under the Marie
Sklodowska-Curie Grant Agreement No. 847523 'INTERACTIONS'. A.G. acknowledges funding from the Hellenic Foundation for Research and Innovation (HFRI) project "4MOVE-U" grant agreement 2688, which is part of the programme "2nd Call for HFRI Research Projects to support Faculty Members and Researchers". L.C. acknowledges support from NSF grant PHY-2308944, NASA ATP award 80NSSC22K0667, and NASA ATP award 80NSSC24K1230. L.S. acknowledges support from DoE Early Career Award DE-SC0023015 and NASA ATP 80NSSC24K1238. This work was also supported by a grant from the Simons Foundation (MP-SCMPS-00001470) to L.S., and facilitated by Multimessenger Plasma Physics Center (MPPC, NSF PHY-2206609 to L.S.).
\end{acknowledgements}

\bibliographystyle{aa}
\bibliography{bibliography}
\begin{appendix}
\section{Synchrotron cooling of muons, pions and kaons}\label{appA}
\begin{figure}[h]
    \centering
    \includegraphics[width=0.8\linewidth]{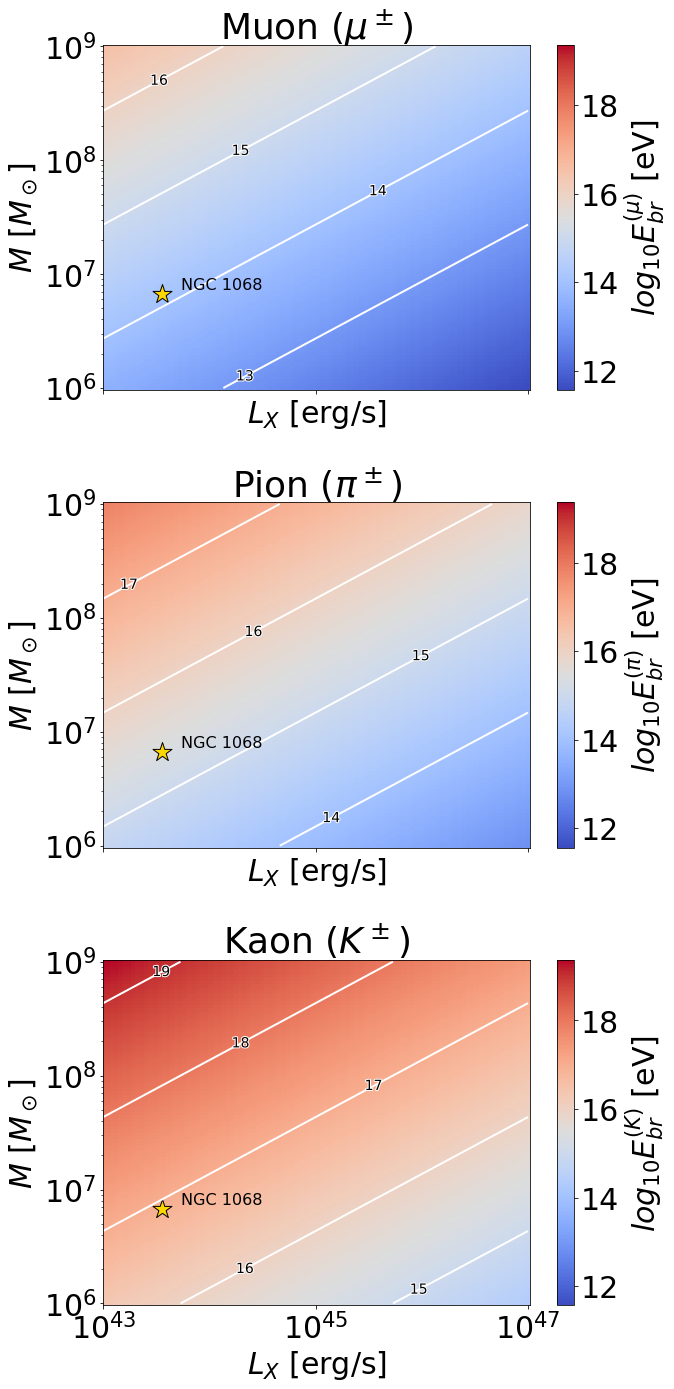}
    \caption{Characteristic energy where the synchrotron energy loss timescale becomes equal to the decay timescale for muons, pions, and charged kaons. The star marks the position of NGC 1068.}
    \label{fig:heatmap}
\end{figure}

\begin{figure}
    \centering
    \includegraphics[width=0.9\linewidth]{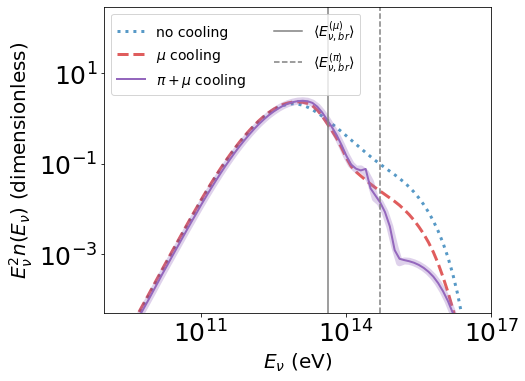}
    \caption{Neutrino spectra displaying the effect of pion and muon cooling. The blue dotted line shows the spectrum without synchrotron cooling while the red dashed and purple solid lines demonstrate the neutrino spectra when only muon cooling or both muon and pion cooling are included, respectively. The thick purple line shows the spectrum after applying a Savitzky-Golay filter to smooth out the numerical artifact appearing at $\sim 10^{14}$~eV. The solid and dashed vertical lines show the position of the cooling break for each species according to Eq.~\ref{eq:Ebr}. }
    \label{fig:meson_cooling}
\end{figure}

Neutrinos are produced by the decay of muons ($\mu^{\pm}$), pions ($\pi^{\pm}$), and kaons ($K^\pm, K^0$). In regions with strong magnetic fields, the synchrotron energy loss timescale can be shorter than the particle decay timescale, causing suppression (damping) of neutrino production above a characteristic energy. For a particle species $i$ with rest mass $m_i$ and decay timescale $\tau_i$ (defined in the particle rest frame) this critical energy is
\begin{equation} 
E_{\rm br}^{(i)} = \sqrt{\frac{3 R}{4 \ell_B c \tau_i}} \left(\frac{m_i}{m_e}\right)^{3/2} m_i c^2,
\label{eq:Ebr}
\end{equation}
where $\ell_B = \sigma_{\rm T} R B^2 / (8 \pi m_e c^2)$ is the magnetic compactness of the corona, a dimensionless measure of its magnetic energy density. 
Figure~\ref{fig:heatmap} shows the corresponding break energies for muons, pions, and charged kaons in coronae with X-ray luminosities between $10^{43}$ erg s$^{-1}$ and $10^{47}$ erg s$^{-1}$, formed around black holes with masses between $10^7 \,M_\odot$ and $10^9 \,M_\odot$ and having a current sheet dimension $L = 6 \,  r_{\rm g}$.  In our coronal model, the magnetic field strength scales as $B \propto L_{\rm X}^{1/2}/R \propto L_{\rm X}^{1/2}/M$, depending on both luminosity and black hole mass. For the parameter space considered here, synchrotron cooling of muons and pions steepens the neutrino spectrum above the corresponding break energies, which lie between 10 TeV and 10 PeV (see also Fig.~\ref{fig:meson_cooling}). 

To generate neutrino spectral templates, we used the numerical leptohadronic code ATHE$\nu$A \citep{2012A&A...546A.120D}. The code solves, over time, a coupled system of integro-differential equations for five particle populations—protons, neutrons, electrons/positrons, photons, and neutrinos—within a fixed volume, through a network of physical processes \citep[see][for a detailed summary]{2024arXiv241114218C}. Synchrotron cooling of unstable particles ($\mu^\pm, \pi^\pm, K^\pm$) is incorporated through an approximate treatment that avoids solving additional equations, as described in \citet{2014MNRAS.445..570P}.

In this scheme, the synchrotron energy lost by each particle prior to decay is first estimated. The remaining energy is then transferred to the secondary particles produced in the decay, with their yields (production rates) precomputed using the Monte Carlo code {\tt SOPHIA} \citep{2000CoPhC.124..290M} \footnote{For an updated discussion on photohadronic cross sections, see \cite{2025arXiv250510674S}.}. The procedure accounts sequentially for synchrotron cooling of $K^\pm$ (whose decays may produce charged pions or muons), then $\pi^\pm$ (which decay into muons), and finally muons themselves. Neutral kaons ($K^0_S$, $K^0_L$) are instead assumed to decay instantaneously, as are neutral pions ($\pi^0$), with their decay products injected directly into kinetic equations of the relevant species. Notably, long-lived neutral kaons ($K^0_L$) can directly produce neutrinos through semileptonic decay channels ($K^0_L \rightarrow \pi^{\pm} + e^{\mp} + \nu_e$, $K^0_L \rightarrow \pi^{\pm} + \mu^{\mp} + \nu_\mu$)\footnote{\url{https://pdg.lbl.gov/2021/listings/rpp2021-list-K-zero-L.pdf}}. These neutrinos ($\nu_e$, $\nu_\mu$) are unaffected by synchrotron cooling\footnote{If proton synchrotron losses dominate at the highest energies, they can steepen the kaon production spectrum, indirectly impacting the neutrino output.}. As a result, above the characteristic pion cooling break energy, the neutrino spectrum is typically dominated by contributions from $K^0_L$ decays \citep{2008PhRvD..77b3007K}.

\begin{figure*}
    \centering
\includegraphics[width=0.98\linewidth]{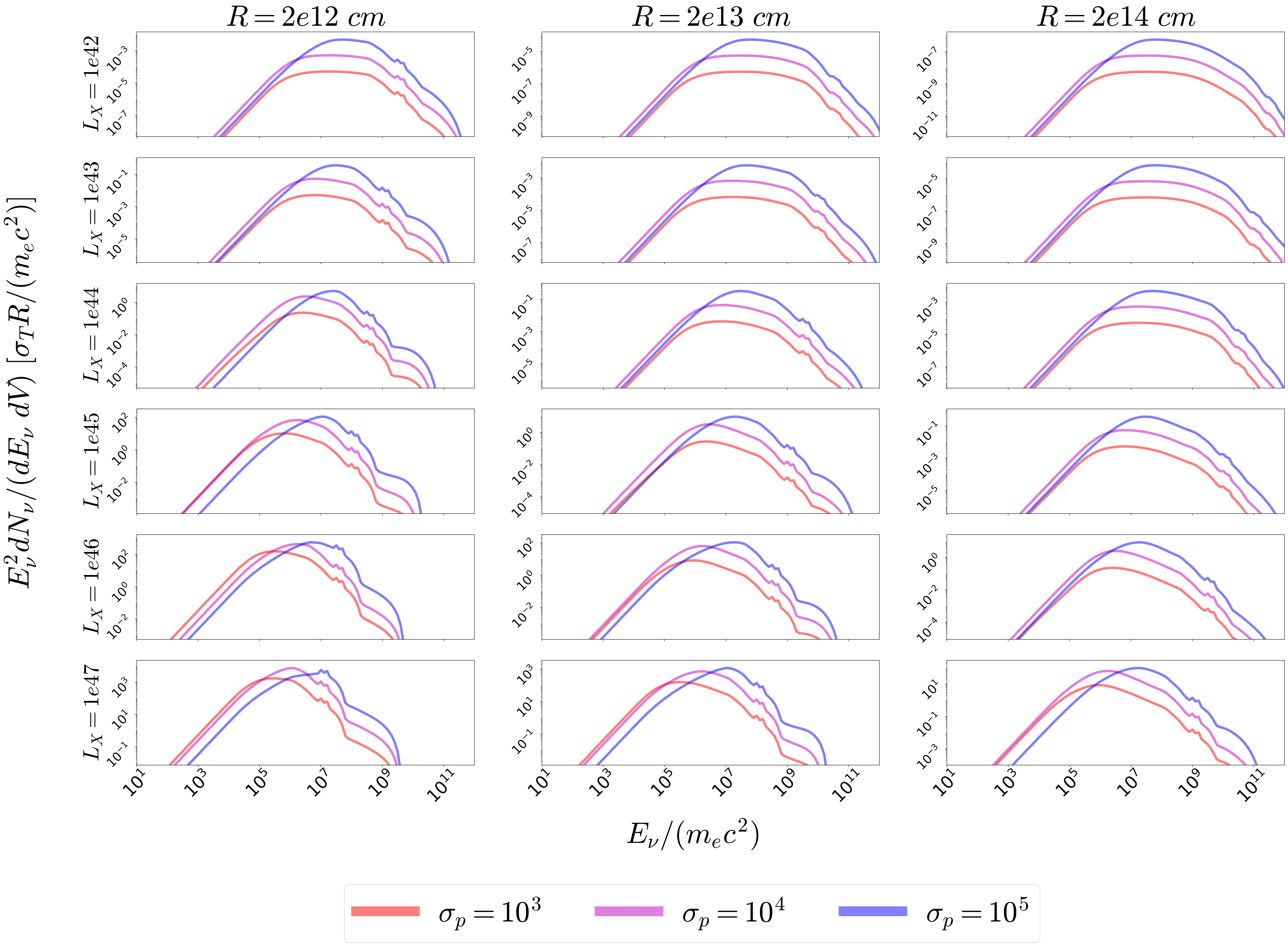}
    \caption{Library of all-flavor neutrino spectral templates computed using the code ATHE$\nu$A for a discrete grid of 0.1-100 keV X-ray luminosities (in erg s$^{-1}$) and coronal radii (in cm). The color of the line represents the value of $\sigma_{\rm p}$ with red, magenta and blue being $10^3$, $10^4$ and $10^5$ respectively.}
    \label{fig:templates}
\end{figure*}

To illustrate more clearly the effects of synchrotron cooling on the neutrino emission we show in Fig.~\ref{fig:meson_cooling} spectra calculated for parameters similar to those of NGC 1068: $L_{\rm X} \simeq 3.6 \times 10^{43}$ erg~ s$^{-1}$, $M \simeq 6.7\times 10^6 M_\odot$,  $B \sim 9 \times 10^4$~G and $\sigma_{\rm p}=10^5$ \citep{karavola_neutrino_2025}. The blue dotted line shows the spectrum without pion and muon cooling while the red dashed and purple solid lines demonstrate the neutrino spectra when only muon or both muon and pion cooling are included, respectively. Inclusion of charged kaon cooling does not alter the neutrino spectrum further. The solid and dashed vertical lines indicate the locations of the cooling breaks for muons and pions, respectively. The sharp feature that appears near the pion cooling energy is a numerical artifact; it is smoothed out when calculating the diffuse flux. Neutrino emission above the peak energy of the spectrum is suppressed by synchrotron cooling of pions and muons, washing out the dependence on the post-break slope of the proton spectrum \citep[for a discussion on the fields required to affect the peak of the NGC 1068 neutrino signal see][]{2025arXiv250915421B}. The high-energy bump that is visible in the purple curve is related to the $K^0_L$ decay. It is also worth noting that the suppression of neutrino flux due to synchrotron cooling of pions and muons weakens as the compactness of the corona $L_{\rm X}/R$ decreases, as illustrated in Fig.~\ref{fig:templates}. 

Figure \ref{fig:templates} presents the neutrino spectral templates calculated for a discrete grid of X-ray luminosity $L_{\rm X}$ and coronal radius $R$, accounting for the cooling of pions and muons. In each panel, we present the all-flavor neutrino spectra (in code units) computed for three different values of the proton magnetization. The suppression of neutrino emission due to synchrotron cooling of pions and muons becomes more pronounced for sources with higher X-ray luminosities and smaller radii. Notably, the neutrino spectra exhibit an approximate self-similarity with respect to $\lambda_{\rm X, Edd}$, as can be seen by comparing panels along the same diagonal. This self-similarity, however, is not exact when synchrotron cooling of pions and muons becomes significant, since the corresponding characteristic break energies (defined in Eq.~\ref{eq:Ebr}) depend on additional parameters beyond $\lambda_{\rm X, Edd}$.

\section{Slopes for protons and X-rays}\label{App:Slopes}

\begin{figure}
    \centering
\includegraphics[width=0.99\linewidth]{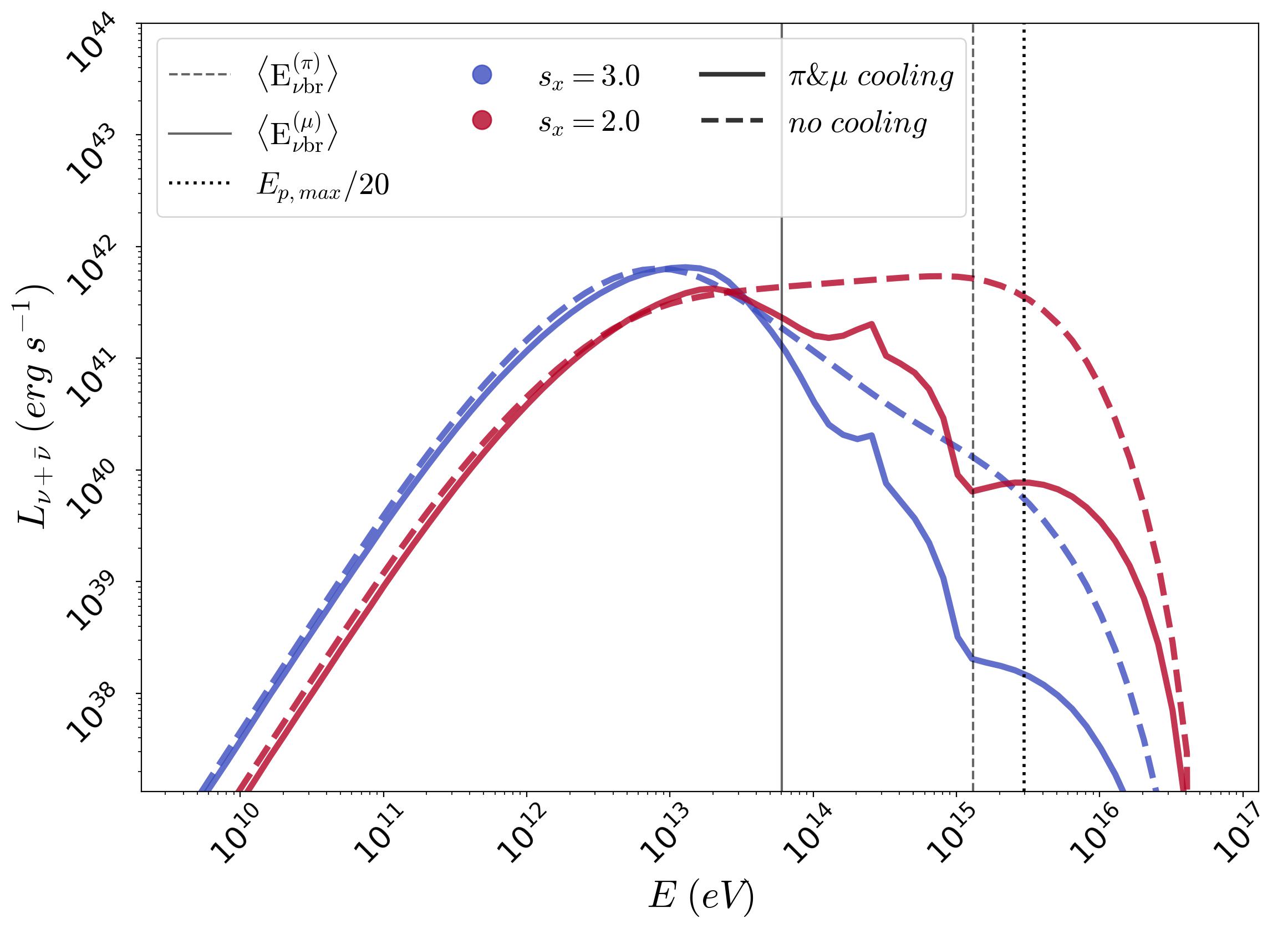}
    \caption{Neutrino spectra displaying the effect of the proton distribution post-break slope $s_{\rm p}$ for the NGC~1068 case-- see Appendix \ref{appA}. Red and blue refer to $s_{\rm p}=2$ and 3 respectively. Solid lines include $\pi$ and $\mu$ cooling while dashed lines neglect it. The solid and dashed vertical lines show the position of the cooling break for each species according to Eq.~\ref{eq:Ebr}. The dotted line refers to the neutrino energy produced by the most energetic protons of the distribution.}
    \label{fig:s_p_effect}
\end{figure}

In our analysis, we adopted $s_{\rm p}=3$ for the post-break proton distribution slope, motivated by prior modeling of NGC~1068 \citep{fiorillo_reconnection_2024}. This slope is controlled by the guide-field strength, $B_{\rm g}$, with $s_{\rm p}=2$ in the limit of vanishing guide fields and $s_{\rm p}=3$ for strong fields (i.e., $B_{\rm g} \sim B_0$ where $B_0$ is the reconnecting magnetic field component \cite{Werner_2017, werner_electron_2024, comisso_pitch-angle_2023}). In figure \ref{fig:s_p_effect} we show the imprint of $s_{\rm p}$ on the produced neutrino spectrum. Solid lines take into account the cooling of heavier particles, as described in Appendix~\ref{appA}, whereas dashed lines ignore the effect. For $s_{\rm p}=2$, in the absence of cooling, the neutrino spectrum is drastically different than for $s_{\rm p}=3$, and its peak is dictated by the maximum energy of the protons rather than $\sigma_{\rm p}$ (compare red and blue dashed lines; see also App. C in \cite{karavola_neutrino_2025}). However, when pion and muon cooling are included, the high-energy ($\gtrsim 30$~TeV) neutrino spectrum softens, and the peak neutrino energy as well as the high-energy slope of the spectrum are essentially determined by the magnetic field. 

\begin{figure}
    \centering    \includegraphics[width=0.99\linewidth]{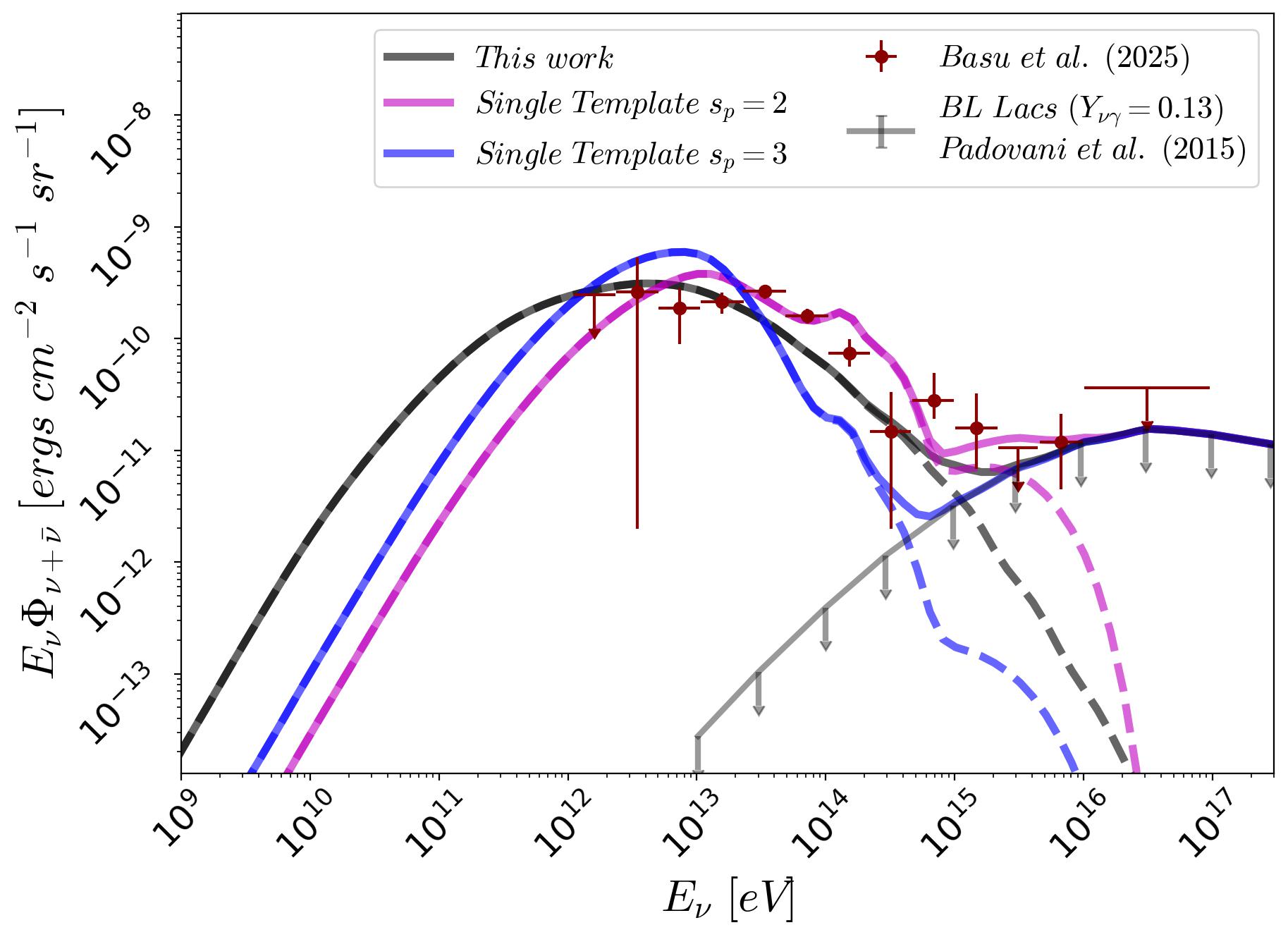}
    \caption{Same as Fig. \ref{fig:diffuse_neutrino_spec}. Magenta and blue lines correspond to the calculation of the diffuse flux assuming that all the sources have the same spectral shape as NGC~1068 (for $s_{\rm p}=2$ and 3 respectively -- see fig. \ref{fig:s_p_effect}) re-normalized in respect to $L_{\rm X}$ and the p$\gamma$ efficiency. Dashed lines refer to the coronal model itself while solid lines also include the BL Lac population.}
    \label{fig:single_template}
\end{figure}

To demonstrate the effect $s_{\rm p}$ has on the diffuse neutrino flux, we performed a simple calculation in which we assigned the NGC~1068 spectral template (shown in fig. \ref{fig:s_p_effect}) to all galaxies after re-scaling it according to $L_{\rm X}$ and the p$\gamma$ efficiency (see \cite{karavola_neutrino_2025} for analytical estimates). We perform this calculation for both $s_{\rm p}=2$ and 3 -- see magenta and blue lines in figure \ref{fig:single_template} respectively. At PeV energies, we clearly see that the single-template analysis with $s_{\rm p}=3$ does not coincide with our calculation, because sources with lower magnetic fields than NGC~1068 dominate in that regime -- see also Appendix~\ref{appC}. As a result, if we were to adopt $s_{\rm p}=2$ for all sources, we would expect the diffuse flux to surpass the observations. Given that the single-template analysis is not representative, to fully account for the effects of the post-break slope, one would have to repeat our analysis for varying $s_{\rm p}$, and we defer this to a future study.

\begin{figure}
    \centering
\includegraphics[width=0.8\linewidth]{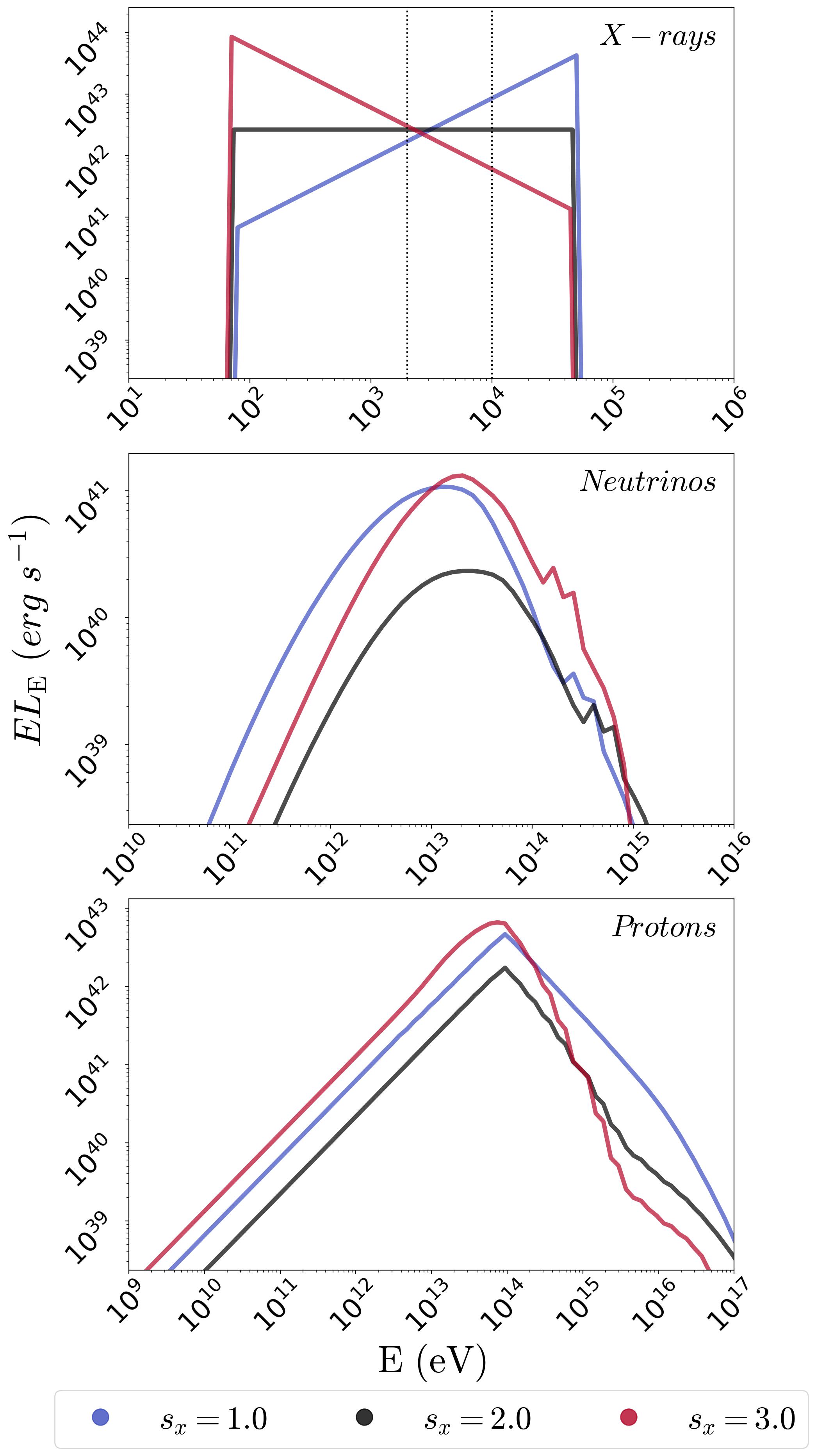}
    \caption{Luminosity distributions for X-rays (top), neutrinos (middle) and protons (bottom) for 3 values of the photon index $s_{\rm x}$--see legend.}
    \label{fig:s_x_effect}
\end{figure}

Furthermore, \cite{trakhtenbrot_bat_2017} showed that the X-ray photon index of non-jetted AGNs is not strongly dependent on the bolometric X-ray luminosity in the 2-10 keV band with its mean value being $\sim 2$. Motivated by the latter result, for the scope of this work, we have assumed $s_{\rm x}=2$. In order to demonstrate the effect of the photon index we show, in figure \ref{fig:s_x_effect}, the distributions of X-rays (top panel), neutrinos (middle panel) and protons (bottom panel) for a single source with $s_{\rm x}=$ 1, 2, and 3. For all three cases, we have used representative parameters close to NGC~1068-- see Appendix \ref{appA}-- for the black hole mass $M$ and proton magnetization $\sigma_{\rm p}$. However, to calculate the bolometric X-ray luminosity $L_{\rm X}|_{s_{\rm x}}$, we considered fixed the luminosity in the 2-10 keV energy band, since it is the value provided by the mock catalog we have used (see Appendix \ref{appB}). We see that for $s_{\rm x}=$ 1 and 3 (red and blue lines), the neutrino spectra are more luminous by a factor of $\sim 5$ compared to the case of $s_{\rm x}=2$ (black line), mostly due to the fact that $L_{\rm p} \propto L_{\rm X}$ is higher in these cases in comparison to the $s_{\rm x}=2$ one. As a result, if all of the sources were described by $s_{\rm x}$=1 or 3, then we would expect the diffuse flux to be higher by a factor of 5, as well.
Moreover, it is interesting to note that high-energy protons ($E \gtrsim \sigma_{\rm p} m_{\rm p} c^2$) appear to \enquote{cool down} more as $s_{\rm x}$ increases. The latter is due to the fact that for higher photon indices there are more photons toward the lower end of the distribution and thus, the efficiency of closer-to-threshold photohadronic interactions is increased.

\section{AGN mock catalogs}\label{appB}

The galaxy and AGN mock catalogs used in this work are constructed using empirical relations on the incidence of accretion events in galaxies \citep[e.g.,][]{Georgakakis2017, Aird2018}. The key quantity in this calculation is the specific accretion rate defined as $\lambda_{\rm SAR}\propto L_{\rm AGN}/M_{\star}$, where $L_{AGN}$ is a proxy of  AGN accretion luminosity and $M_{\star}$ is the stellar mass of the host galaxy. The $\lambda_{\rm SAR}$ measures the accretion luminosity produced per unit stellar mass of the host galaxy and under certain assumptions, namely bolometric correction and scaling relation between stellar mass and black hole mass, it can be converted to the Eddington ratio, $\lambda_{\rm Edd}$.  For the purpose of our analysis, a convenient feature of $\lambda_{\rm SAR}$ is that it can be measured observationally for large and unbiased samples of AGN. This is unlike the Eddington ratio that requires a direct estimate of the black hole mass. This latter step is observationally challenging as it relies on independent methods, such as measurement of broad optical emission lines or accretion disk SED fitting \citep[e.g.,][]{2014SSRv..183..253P, 2018A&A...618A.127V, popovic_2025, 2025arXiv250906933K}. For certain populations, such as obscured or low-luminosity AGN with no detectable broad emission lines, black hole mass measurements turn out to be impossible.

Extragalactic survey fields that benefit from observations in many different wavebands, from X-ray to the far-infrared, have made possible the compilation of large and nearly unbiased AGN samples, for example via detection at X-rays, out to high redshift and the estimation of their host galaxies properties, such as stellar masses. Such observations allow for the measurement of the fraction of galaxies in a given stellar mass interval that host AGN with a specific accretion rate $\lambda_{\rm SAR}$. These fractions can then be turned into probability distribution functions, $P(\lambda_{\rm SAR})$, which describe the probability of accretion events with $\lambda_{\rm SAR}$ in galaxies  \citep[e.g.,][]{Georgakakis2017, Aird2018}. The  functional form of $P(\lambda_{\rm SAR})$ can be approximated by a broken power-law with parameters that evolve with cosmic time. By construction, the  $P(\lambda_{\rm SAR})$ has the feature that if convolved with the stellar mass function of galaxies, it yields the AGN luminosity function

\begin{equation}
    \Phi(L_{\rm AGN}, z) = \int \psi(M_\star, z)\cdot P[\lambda_{SAR}(L_{\rm AGN}, M_\star), z]\cdot {\mathrm d}\log M_{\star},
\end{equation}

\noindent where $\Phi(L_{\rm AGN}, z)$ and $\psi(M_\star, z)$ are the AGN luminosity function and the galaxy stellar mass function, respectively, at redshift $z$ and the specific accretion rate is expressed as a function of $L_{\rm AGN}$ and $M_\star$. It is exactly this feature that we exploit in this work to populate mock galaxies randomly drawn from a stellar mass function with AGN. 

In practice, parametric forms for $\Phi(L_{\rm AGN}, z)$ and $\psi(M_\star, z)$ are assumed to solve the convolution equation above and estimate $P(\lambda_{\rm SAR}, z)$. We can then draw galaxy samples from the stellar mass function and paint AGN on them by probabilistically assigning them specific accretion rates from the empirically determined $P(\lambda_{\rm SAR}, z)$. The end product of this approach is a catalog of mock galaxies, each of which is associated with a redshift, stellar mass, $\lambda_{\rm SAR}$ and hence an AGN luminosity $L_{\rm AGN}\propto\lambda_{\rm SAR}\cdot M_\star$. The last step is to assign each galaxy a black hole mass (and therefore an Eddington ratio given the AGN luminosity) using observationally inferred scaling relations between black hole and stellar masses. 

We implement the above methods as described in \cite{georgakakis_forward_2020}. We adopt the X-ray AGN luminosity function of \cite{2014ApJ...786..104U} and the parametric stellar mass functions of \cite{Ilbert2013}. The specific accretion rate probability distribution function follows a three-segment broken power-law, motivated by the results of non-parametric studies \citep[e.g.,][]{Georgakakis2017, Aird2018}. For the black-hole versus stellar mass correlation we adopt the one of \cite{marconi_relation_2003}, $M_{\rm BH}=2 \cdot 10^{-3} M_{*}$. Experimentation with other scaling relations in the literature shows that our results are not sensitive (factor of $<2$) to the adopted relation between $M_{\rm  BH}$ and $M_\star$.  The final mock catalog includes for each galaxy the: stellar mass, redshift, specific accretion rate, black hole mass, and 2-10\,keV X-ray luminosity of the AGN. 

Since the generation of the mock catalog includes a sampling step for the stellar mass function, one has to assume a volume that the galaxies occupy, which translates to a fiducial sky area of the mock survey. The baseline mock catalog, which extends up to $z=4$, corresponds to an area of $50\,\rm deg^2$. However, for this solid angle the sampled volume at low redshift, $z\la0.2$, is too small and our results are affected by poor statistics. We therefore choose to complement the baseline mock with a low redshift one with a solid angle of $\rm 20,000\,deg^2$ and limited to $z\le0.2$. For our analysis, we use the low redshift mock for $z\le0.2$ and the baseline one for higher redshifts.  

The modeling approach above has the advantage of producing realistic mock AGN catalogs that include host galaxy properties (e.g., stellar mass, black hole mass) and by construction reproduce fundamental statistics of the AGN population, such as space densities, host galaxy stellar mass function, large scale structure, multiwavelength characteristics \citep[see][for details]{georgakakis_forward_2020}.

\begin{figure}
    \centering    \includegraphics[width=0.99\linewidth]{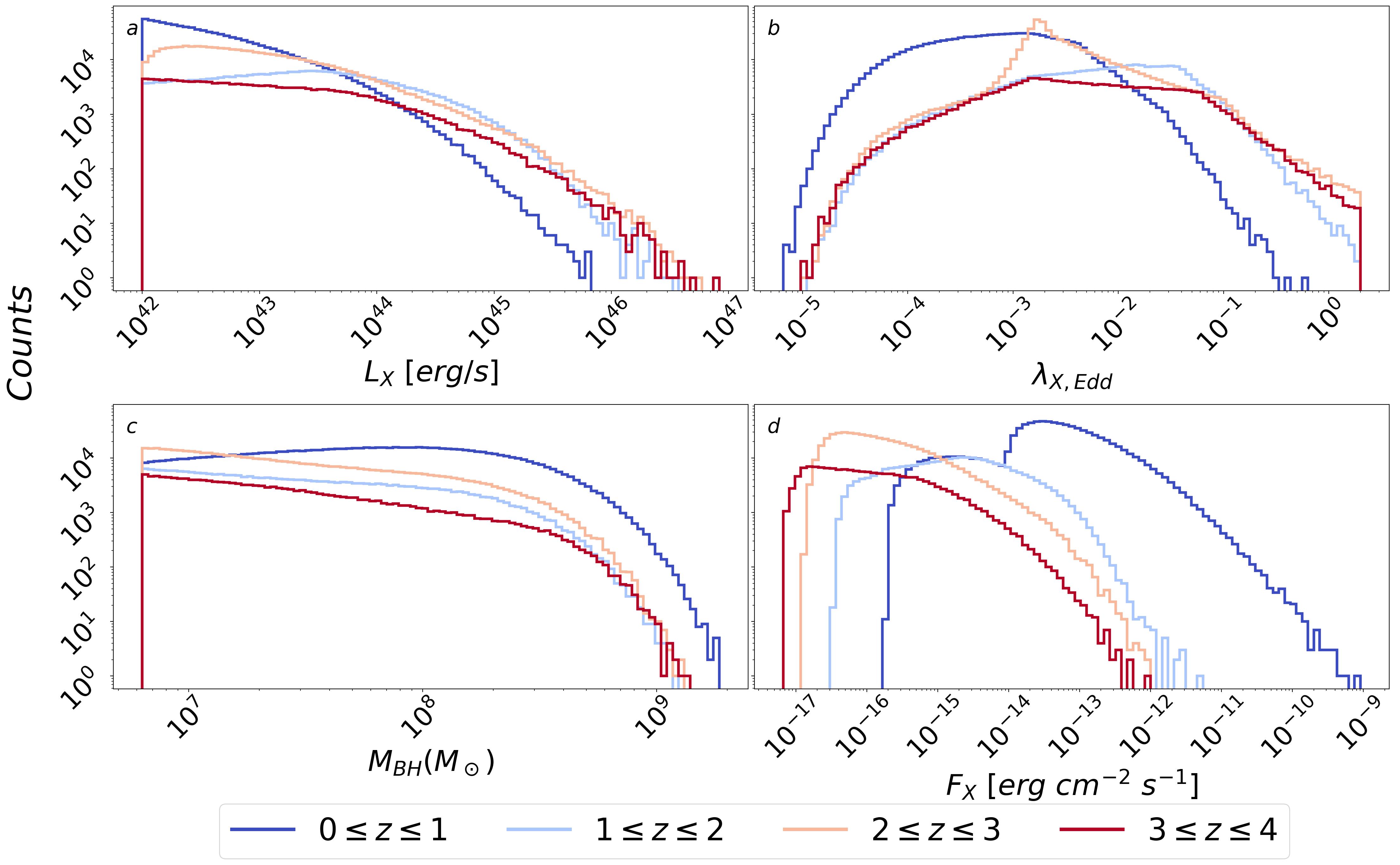}
    \caption{Histograms displaying properties of AGN from the mock catalog: (a) bolometric X-ray luminosity $L_{\rm X}>10^{42}$~erg s$^{-1}$, (b) X-ray Eddington ratio $\lambda_{\rm X, Edd}$, (c) black hole mass, and (d) the bolometric X-ray flux. Color shows different redshift bands, see legend.}
    \label{fig:mock}
\end{figure}

Figure \ref{fig:mock} shows the distribution of different properties of the mock galaxies: (a) bolometric ($0.1-100$~keV) X-ray luminosity, (b) X-ray Eddington ratio (defined as $L_{\rm X}/L_{\rm Edd}$), (c) black hole mass, and (d) bolometric X-ray flux. For each panel, different colors correspond to different redshift bands (see legend). The mock catalog contains AGN with black hole masses from $\sim 10^7 M_\odot$ up to $\sim 10^9 M_\odot$ with a distribution that does not depend on their distance, as seen in panel c. AGN with high bolometric X-ray luminosities ($L_{\rm X} \gtrsim 10^{45}$~erg s$^{-1}$) are mostly found at large distances ($z \gtrsim 1$), while AGN with higher X-ray fluxes can be found relatively closer to us ($z \lesssim 1$).

\section{Decomposition of the diffuse neutrino flux}\label{appC}

\begin{figure}[h]
    \centering
    \includegraphics[width=0.99\linewidth]{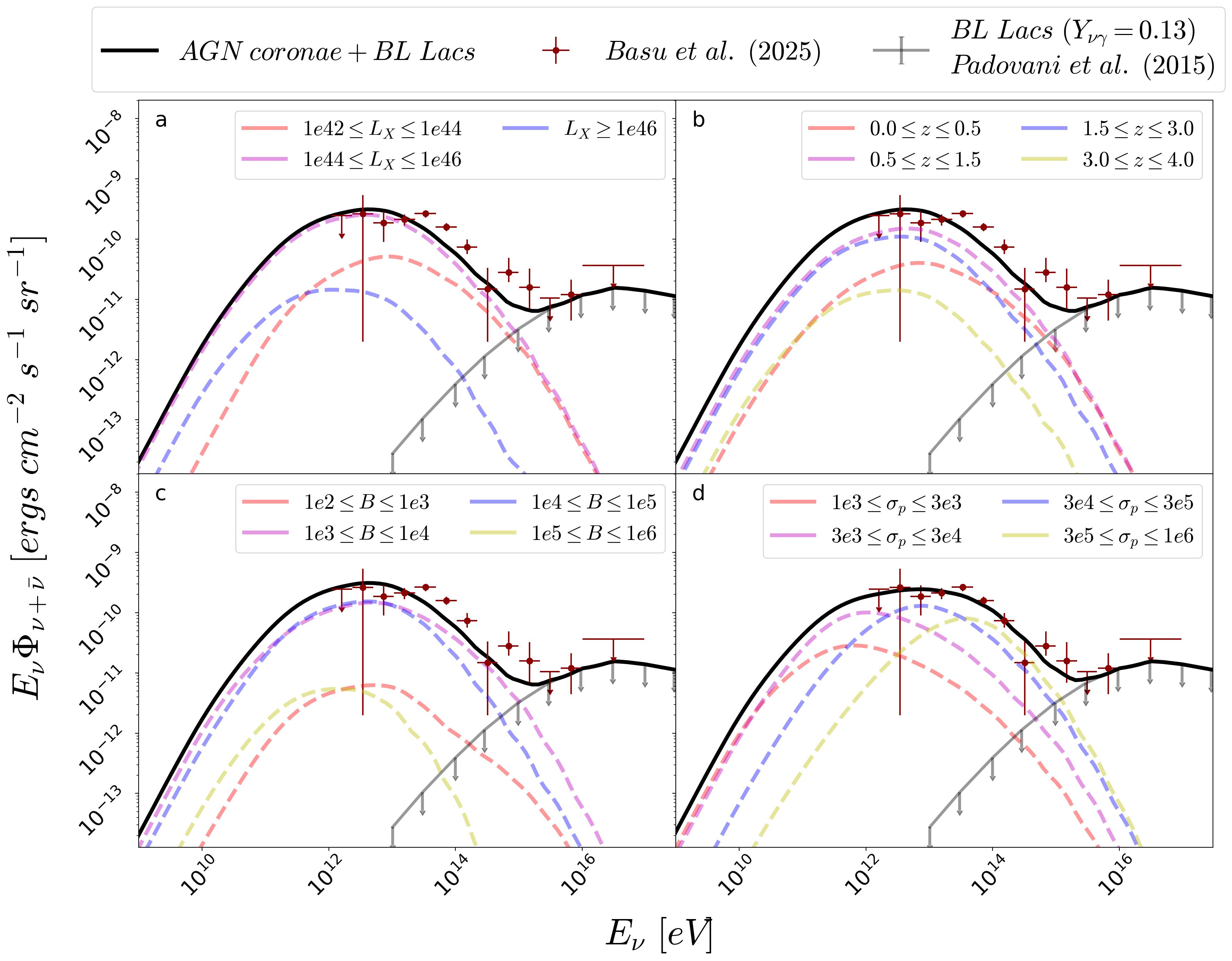}
    \caption{Same as Fig. \ref{fig:diffuse_neutrino_spec} except that lines are color coded based on the bolometric X-ray luminosity (panel a), the redshift z (panel b) and on the magnetic field (panel c) of AGN as indicated by the inset legend. Panel d shows the exact same calculation as in Fig. \ref{fig:diffuse_neutrino_spec} but for $\sigma_{\rm p}$ extending up to $10^6$.}
    \label{fig:LX_conrtib}
\end{figure}

In our model, the neutrino luminosity of a corona that is optically thin to p$\gamma$ interactions scales linearly with the proton plasma magnetization \citep{karavola_neutrino_2025}. This has an imprint on the diffuse flux as demonstrated in Fig.~\ref{fig:diffuse_neutrino_spec}. Another key parameter of our model, which controls the expected neutrino luminosity, is the bolometric X-ray luminosity of the corona (see Eq.~3.6 in \cite{karavola_neutrino_2025}). Moreover, the redshift of the source affects the observed neutrino flux. In Fig.~\ref{fig:LX_conrtib} we show the contribution of different X-ray luminosity bins (panel a), redshift bins (panel b) and magnetic field bins (panel c) to the total neutrino flux. Solid lines are the same as in Fig. \ref{fig:diffuse_neutrino_spec}, with colors representing the range of X-ray luminosities, redshifts or magnetic fields. One can see that the dominant contribution originates from galaxies at $0.5 \leq z \leq 3.0$ with $10^{44} \, {\rm erg~s}^{-1} \leq L_{\rm X} \leq 10^{46} \, {\rm erg~s}^{-1}$ and $10^3~ \rm G\leq B \leq 10^5~G$.

Furthermore, panel d in Fig.~\ref{fig:LX_conrtib} shows the same calculation as in Fig. \ref{fig:diffuse_neutrino_spec} but with $\sigma_{\rm p}$ extending up to $10^6$. We note that this solution can better describe the observations in the PeV energy range. However, a softer power-law distribution with $n=4.5$ is needed to avoid overshooting the data. Nonetheless, the take-home message remains the same: only a small fraction of the non-jetted AGN population should have high proton plasma magnetizations. In particular, $\sim15\%$ of the sources should have $\sigma_{\rm p} \geq 5 \times 10^4$ in order to match the diffuse flux observations. In this case, coronae with magnetizations higher than the one describing NGC~1068 constitute a very small fraction of approximately $4 \%$ of the population. As a result, the detection probability of such high-$\sigma_{\rm p}$ coronae is expected to be low.

\begin{figure}
    \centering
    \includegraphics[width=0.9\linewidth]{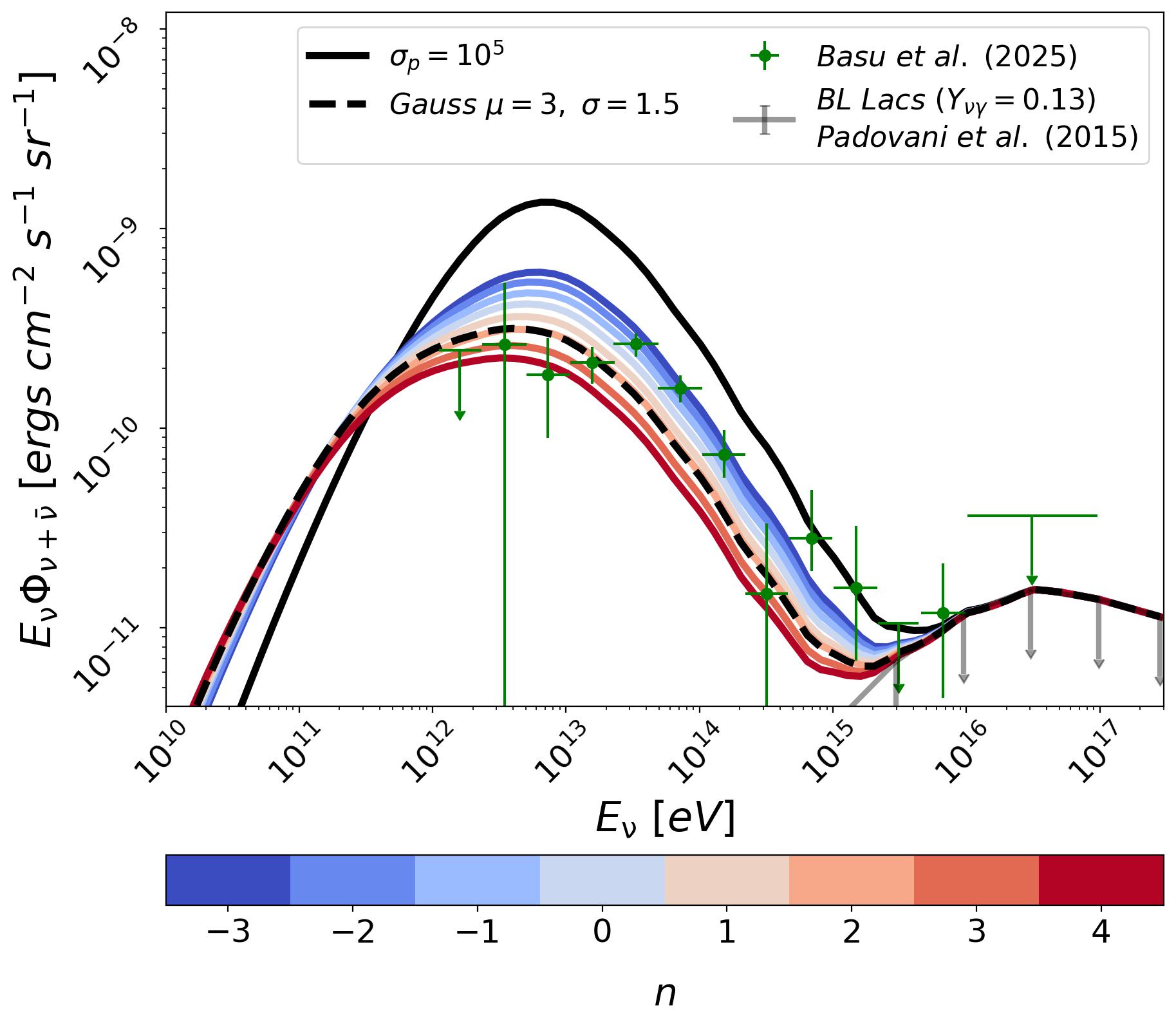}

    \caption{Similar as Fig. \ref{fig:diffuse_neutrino_spec} except that black solid line shows result of the assumption that all AGN have $\sigma_{\rm p}=10^5$, while colored lines occur from power-law probability distributions of different slopes n (see colorbar) and black dashed line comes from $P(\sigma_{\rm p})$ being a gaussian distribution with a mean value $\mu=3$ and a standard deviation of $\sigma=1.5$.}
    \label{fig:slopes}
\end{figure}

Finally, Fig. \ref{fig:slopes} shows the impact of different probability functions $P(\sigma_{\rm p})$ on the diffuse flux calculation. Black solid line assumes $\sigma_{\rm p}=10^5$ for all sources of the catalog, while colored lines show results for $P(\sigma_{\rm p})\propto \sigma_{\rm p}^{-n}$ with the color indicating the value of the slope $n$ (see color bar). Additionally,  the black dashed line is computed using a Gaussian distribution with respect to $\log_{10}(\sigma_{\rm p})$ with mean value $\mu=3$ and standard deviation $\sigma=1.5$ so that, again, 10\% of the non-jetted AGNs have a value of $\sigma_{\rm p} \sim 10^5$. If we attribute $\sigma_{\rm p}=10^5$ to all galaxies, the diffuse flux overshoots the data by almost one order of magnitude at $E_\nu \sim 10$~TeV. Also, we note that changes in the slope $n$ of the $\sigma_{\rm p}$ probability distribution weakly affect the diffuse flux calculation, with $1<n<3$ resulting in the best description of the IceCube measurements \cite{basu_measurement_2025}. Interestingly, adopting a Gaussian probability distribution leads to very similar neutrino spectra as those obtained for the power-law distributions. We conclude that the shape of $P(\sigma_{\rm p})$ does not have a large impact on the results as long as the fraction of galaxies with $5\times 10^4\lesssim \sigma_{\rm p}$ is about $10\%$.

\end{appendix}

\end{document}